# Charge Conjugation and Parity Violations as a Signature for Black Hole Formation in Hadron Collisions

Michael J. Longo

Department of Physics, University of Michigan, Ann Arbor, MI 48109-1120 USA

Abstract

Black hole formation (BHF) in hadron collisions is a subject of great interest. Hard quark-quark collisions are generally associated with the production of particles at large transverse momenta or with high multiplicities. If BHF occurs, we might expect large violations of baryon number, charge conjugation (*C*), parity (*P*), strangeness, isospin, *etc*. in such processes. *C* and *P* have not been well tested in hadronic interactions. I propose new tests of *C* and *P* that can be adapted to a variety of BHF scenarios. These may provide more sensitive searches for BHF or other new physics at existing colliders.

Black hole formation in hadron collisions is a subject of great current interest.[1, 2] Speculation about catastrophic black hole formation at colliders has even appeared in the popular press.[3] Though the possibility of such a spectacular signature as destroying the Earth has largely been disproved,[4] more subtle signals of black hole formation in hadron collisions have proven to be elusive.

Most of the recent discussion of BHF has been in the context of the CERN Large Hadron Collider, though potentially observable rates could occur at the Fermilab Tevatron if the fundamental gravity scale is below 1.4 TeV.[5] The proposed signatures for BHF range from lepton production at large transverse momenta[2] to the presence[6] (or absence[7]) of multijet events with large transverse momenta, as well as the presence[1] (or absence[8]) of large missing energy. The formation of a quark-gluon plasma decaying into thousands of photons has also been suggested[9]. No experimental evidence of BHF has yet been found, but it is important to explore all possible signatures.

It has long been believed that only quantum numbers and symmetries that are associated with long-range fields are preserved in black hole formation.[10] Thus charge, angular momentum, and mass are presumed to be conserved, while baryon number and lepton number, as well as charge conjugation, parity, isospin, strangeness and other quantum numbers that are believed to be conserved only in hadronic interactions, can be expected to show large violations if BHF and subsequent decay by Hawking radiation[11] occurs in hadron collisions. Here I discuss how such signatures of black hole formation might manifest themselves and how they might best be observed (or perhaps might already have been observed).

Black hole formation is presumed to occur in the most violent quark-quark collisions. Thus the most promising venue for finding BHF before the CERN LHC is the Fermilab Collider. Because hard quark-quark collisions are generally associated with the production of particles at large transverse momenta or with high multiplicities, if BHF occurs we might expect large violations of baryon number, charge conjugation, strangeness, isospin, *etc.* in such processes. While violations of baryon or lepton number conservation would be a very dramatic signature, as a practical matter all the existing collider detectors have large uninstrumented regions where most of the particles from interactions are lost. This limitation, as well as the lack of particle species identification except for low momentum particles, makes tests via baryon/lepton number conservation, strangeness, *etc*. impractical.

One promising method to search for BHF appears to be testing charge conjugation (*C*) in hadronic interactions at large momentum transfers. Ironically, there have been very few tests of *C* in strong interactions generally,[12] and none at large momentum transfers or collider energies. This is surprising as a $\bar{p}$-*p* collider with charge conjugate particles of equal momentum in the initial state is an ideal laboratory for testing *C* invariance.

An immediate prediction of *C* invariance is that the spectrum of any charged particle in the forward $\bar{p}$ hemisphere should be identical to that of the charge conjugate particle in the forward *p* hemisphere. This can be tested in a collider detector with a magnetic field, even without particle identification. The only relevant data appear to be from the CDF detector. These unpublished data from the thesis of A. Byon are reproduced in Fig. 1. In this figure the pseudorapidity $\eta \equiv -log_e(tan\frac{\theta}{2})$, so that $\eta = 0$ corresponds to 90° with respect to either beam. Thus the expectation of charge conjugation is that the positive and negative distributions should be mirror symmetric about $\eta = 0$. Unfortunately, the numerical data are no longer available, but a careful evaluation of the graphed data shows no sign of any *C*-violating asymmetry. However, it should be noted that these data are at low transverse momentum, $\langle p_t \rangle \approx 1$ GeV/c, so are not a good test for BHF. A careful test should be done with high $p_T$ particles and would require corrections for apparent $C$ violations due to weak decays, such as $\Lambda \rightarrow p + \pi^-$, which generally show large $C$ violations.

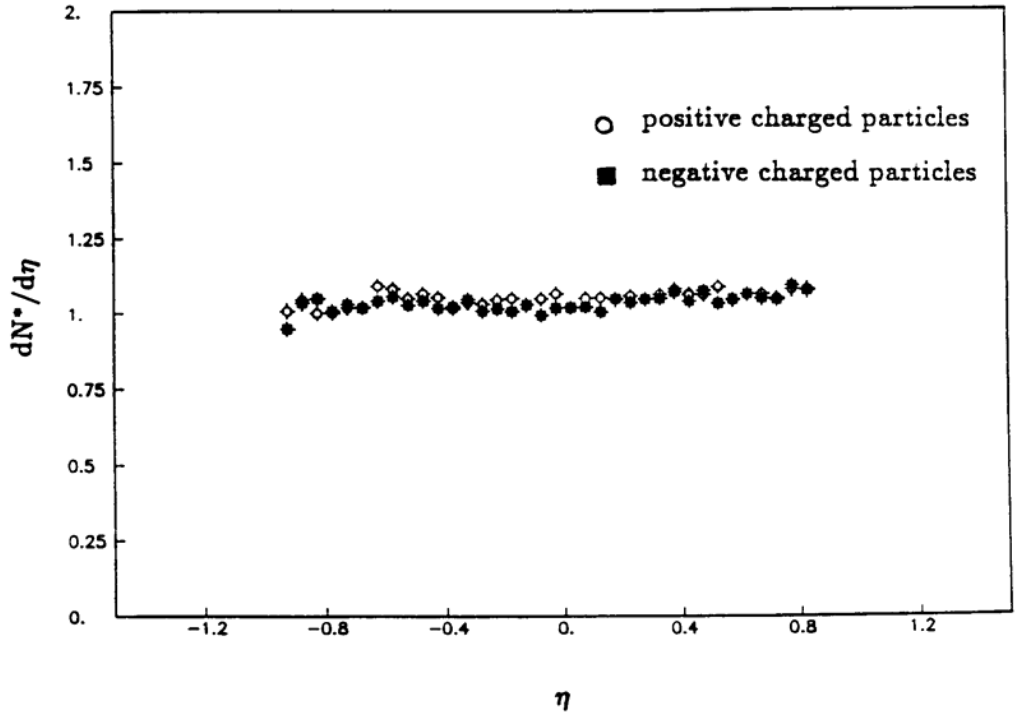

**Figure 1 Distribution of positive and negative particles from central collisions in the CDF detector from the thesis of A. Byon.** $\eta \equiv -log_e(tan\frac{\theta}{2})$ **Charge conjugation requires that they be mirror symmetric about $\eta = 0$.**

Large parity violations might also occur due to BHF. This could be tested by searching for parity-violating spin components of hyperons such as $\sigma \cdot p$ where $p$ is the vector momentum of the hyperon. The large asymmetry in $\Lambda^0$ hyperon decays makes this a particularly attractive possibility for a sensitive search for BHF. In this search one can choose whatever sample of events that are most likely to be associated with BHF, such as high multiplicity, high (or low) transverse momentum, high (or low) missing energy, *etc.* Another possibility is to look for parity-violating correlations such as a term $p_1 \cdot (p_2 \times p_3)$, where $p_1, p_2, p_3$... are any combination of beam particle, lepton, photon, $W^\pm$, $Z^0$, high $p_T$ particle, or jet momenta.[13] Again, a variety of event types and correlations can be looked at to investigate a variety of BHF scenarios.

BHF would also presumably lead to violations of *CP*, *T*, and possibly *CPT*. The mirror symmetry of charged particle production discussed above as a test of *C* is also required by *CP*. Though violations of causality and *CPT* might be considered the holy grail of BHF, it is not obvious how to test these. It is not possible to test for isospin violations due to BHF directly. However, if BHF does occur, there is no constraint on isospin in the final state, so that large fluctuations in the ratio of charged to neutral pions might be expected. Such fluctuations have been observed in cosmic ray events ("Centauro" and "anti-Centauro" events with few $\pi^0$ and mostly $\pi^0$, resp.).[14] These could be a signal for BHF, though many other interpretations have been suggested.[14]

Tests of *C* and *P* at large $p_T$ from electron-positron collisions are also of interest. The apparent absence of electric dipole moments of the electron and neutron[12] are excellent tests of *P* in electromagnetic interactions. Tests of *C* in electromagnetic interactions have been made[12] by searching for *C*-violating decays of low-mass particles such as

$\eta^0 \to \pi^0 + \mu^+ + \mu^-$, but no tests of *C* or *P* have been made at colliders at large transverse energies.

I conclude that violation of discrete symmetries, such as *C* and *P*, may be the best "smoking gun" signature of black hole formation in hadronic interactions. Relatively straightforward experimental tests are possible, very few of which have been done. In any case, these symmetries have never been tested at large transverse momenta, and large violations could occur from black hole formation or other "new physics". Causality and *CPT* violation could also occur, though it is not clear how these would manifest themselves or be tested for. If BHF can be identified and their properties studied through their violations of symmetries and conservation laws, it will provide a long sought connection between quantum mechanics and general relativity. Copious production of black holes is predicted at the LHC; they may already be produced at observable rates at the Fermilab Tevatron if more sensitive tests are done.

---